\documentclass[showpacs,amsmath,amssymb,superscriptaddress]{revtex4}
\usepackage{amsmath,amssymb,color,latexsym,natbib,graphicx}
\usepackage{dcolumn}
\usepackage{bm}
 
\newcommand{\p} {\partial}

\def\uu{{\bf u}}

\def\d_M{{\bf d_M}}

\def\rr{{\bf r}}

\def\xx{{\bf x}}

\def\be{\begin{equation}}
\def\ee{\end{equation}}
\def\ba{\begin{eqnarray}}
\def\ea{\end{eqnarray}}
\def \pmbmath{\mathpalette\pmbmathaux}
\def \pmbmathaux#1#2{
         \pmbtext{$#1#2$}}
\def \pmbtext#1{\leavevmode
     \setbox0\hbox{#1}
     \kern0,4pt \copy0 \kern-\wd0
     \kern-0,2pt \raise0,3pt \box0 }
 
 \usepackage{microtype}
 \usepackage{calrsfs}
 \DeclareMathAlphabet{\pazocal}{OMS}{zplm}{m}{n}
 \usepackage[cal=rsfso,calscaled=.96]{mathalfa}
 \DeclareMathAlphabet{\mathpzc}{OT1}{pzc}{m}{it}
 \DeclareMathAlphabet{\mathcalligra}{T1}{calligra}{m}{n}

 \newcommand{\aap}{Astron. Astrophys.}
 \newcommand{\aapr}{A\&A Rev.}
 \newcommand{\mnras}{Mon. Not. R. Astron. Soc.}
 \newcommand{\apjl}{Astrophys. J. Lett.}
 \newcommand{\jfm}{J. Fluid Mech.}

 \def\uu{{\bm u}}
 \def\gg{{\bm g}}
 
 \def\ag{{\bm {A_G}}}
 \def\jm{{\bm {j}}}
 \def\rr{{\bm r}}
 \def\kk{{\bm k}}
 \def\xx{{\bm x}}
 \def\Nabla{{\pmbmath {\nabla}}}
 \def\be{\begin{equation}}
 \def\ee{\end{equation}}
 \def\ba{\begin{eqnarray}}
 \def\ea{\end{eqnarray}}
\begin{document}

\preprint{1}

\title{{Exact Relations for Energy Transfer in Self-gravitating Isothermal Turbulence}}
\author{Supratik Banerjee}
\email{supratik.banerjee@uni-koeln.de}
\affiliation{Universit\"at zu K\"oln, Institut fur Geophysik und Meteorologie, Pohligstrasse 3, 50969 K\"oln, Germany}
\author{Alexei G. Kritsuk}
\email{akritsuk@ucsd.edu}
\affiliation{Center for Astrophysics and Space Sciences, University of California, San Diego, \\9500 Gilman Drive, La Jolla, California 92093-0424, USA}

\date{\today}

\begin{abstract}
Self-gravitating isothermal supersonic turbulence is analyzed in the asymptotic limit of large Reynolds numbers. Based on the inviscid invariance of total energy, an exact relation is derived for homogeneous, (not necessarily isotropic) turbulence. A modified definition for the two-point energy correlation functions is used to comply with the requirement of detailed energy equipartition in the acoustic limit. In contrast to the previous relations (Galtier and Banerjee, Phys. Rev. Lett., 107, 134501, 2011; Banerjee and Galtier, Phys. Rev. E, 87, 013019, 2013), the current exact relation shows that the pressure dilatation terms plays practically no role in the energy cascade. Both the flux and source terms are written in terms of two-point differences.  Sources enter the relation in a form of mixed second-order structure functions. Unlike kinetic and thermodynamic potential energy, gravitational contribution is absent from the flux term. An estimate shows that for the isotropic case, the correlation between density and gravitational acceleration may play an important role in modifying the energy transfer in self-gravitating turbulence. The exact relation is also written in an alternative form in terms of two-point correlation functions, which is then used to describe scale-by-scale energy budget in spectral space.

\end{abstract}
\maketitle

\section{Introduction}

The formation of stars in giant molecular clouds is one of the most intriguing subjects in astrophysics \citep[e.g.][]{krumholz14}. 
Indeed, the star formation rate is surprisingly low in present-day spirals in the sense that only about $1\%$--$10\%$ of the gas forms stars every 
free-fall time \citep{Zuckerman1974,Krumholz2007,Hennebelle11}. 
How (and if) the global collapse of self-gravitating molecular clouds is prevented is a subject of controversy. 
Many factors, including the hydrostatic pressure, rotation, magnetic field, are likely to play a significant role in this process 
\citep[e.g.][]{McKee07,padoan+14}. However, these clouds exhibit strong density and velocity fluctuations 
which make supersonic turbulence a potential candidate in supporting a self-gravitating cloud against the collapse.
Despite early attention by theorists \citep{Weizsacker51, Chandrasekhar51, Parker52}, the subject of turbulence 
was subsided by the intermediate interest on studying more tractable effects of rotation \citep{Schmitz83, Schmitz84, Schmitz86,Terebey84} and magnetic field \citep{Mouschovias76,Shu77,Shu87}.
It is only relatively recently that a number of observational and numerical evidences indicate that the collapse of a self-gravitating cloud is 
mainly regulated by supersonic turbulence \citep[and references therein]{Leorat90, Pouquet91,pouquet..91,hennebelle.12} 
and so is the mass and the star formation rate.
In this framework, several statistical studies (mainly numerical) have also been performed on supersonic turbulence 
\citep{kritsuk...07,Schmidt09,Federrath10,kritsuk..11,Federrath,kritsuk..13,Schmidt13}. 
The role of turbulence is basically two-fold, i.e. to provide stability at large scales and to initiate collapse at smaller scales by 
local density enhancements \citep{Larson03, Maclow04}. Therefore the presence of turbulent support basically hinders the 
star formation process. Unfortunately, to date, neither any clear analytical explanation nor a satisfactory quantitative estimation exists 
in order to properly understand the corresponding physics: existing analytical works are mostly based on incompressible or weakly compressible 
turbulence \citep{Bonazzola87,Biglari88}, whereas interstellar turbulence can be highly supersonic with turbulent Mach numbers in excess of $10$ \citep[e.g.][]{heyer.04}. It is worth noting that weakly compressible turbulence, which allows a Kolmogorov spectrum, can be shown to 
bring about collapse approximately at the Jeans scale \citep{Sasao71, Sasao73}. 

A number of numerical studies of self-gravitating supersonic turbulence 
ranging from 1D to 3D have been reported 
starting from mid-1990s \citep[e.g.][]{Vazquez95, Vazquez96,Klessen00, padoan...05, Ballesteros07, kritsuk..11, collins......12, Schmidt13,FederrathK2013}. 
Both the early low resolution spectral simulations and more recent high resolution 3D simulations conclude that turbulence generated by the gravity alone (with or without additional support by a magnetic field) is insufficient to slow down significantly the star formation rate, 
and the turbulent regulation against gravitational collapse is nothing but a global effect whereas locally turbulence can promote collapse 
by increasing the density \citep{kritsuk..17}. Current global star formation models invoke an energy feedback loop, effectively reducing the mean gas density in response to elevated star formation activity levels \citep{ostriker..10,hopkins......14,kim.15}. 
As far as the structure of turbulent self-gravitating clouds is concerned, there is no comprehensive theory available to describe it or predict rigorously any sort of energy equipartition or quasi-virialization across scales, 
even though recent observations and numerical simulations seem to indicate that certain patterns are ubiquitously present in the data \citep{murray...15,Stanchev2015,veltchev..16,li.16}.

In this paper, we study analytically the role of self-gravity in fully developed isothermal supersonic turbulence which is a valid model for the star-forming interstellar clouds. Using two-point statistics, we derive an exact relation for homogeneous and statistically stationary turbulence whereas no prior assumption of isotropy is made.  The motivation of this paper is two-fold. The derivation of the exact relation relates to general compressible turbulence whereas the principal physical motivation lies in understanding the role of gravity in cold star-forming clouds where the turbulence is usually supersonic. The derivation follows the methodology previously outlined by
\citet{Galtier11,Banerjee13, Banerjee14}. However, this work revisits the definition of two-point correlation functions which renders the final exact relation simpler and entirely expressible in terms of two-point fluctuations. 
The paper is structured as follows. In section 2 we discuss the energy conservation. The following section explains the contruction of correlation functions in details. 
The derivation of the exact relation is given in section 4 where three alternative forms of the exact relations are discussed. Finally section 5 presents a summary of the paper. 


\section{Energy conservation}
We start our analysis with the following equations describing a self-gravitating compressible fluid in three dimensions
\begin{eqnarray}
\p_t \rho + \pmbmath{\nabla} \cdot (\rho \uu ) &=& 0,  \label{continuity}\;\;\;\;\;\;\;\; \\
\partial_t (\rho \uu )+ \pmbmath{\nabla} \cdot (\rho \uu \otimes \uu) &=& - \pmbmath{\nabla} p + \rho {\gg}  + {\bm d} + {\bm f}, \label{momentum} \\
\nabla \cdot {\gg} &=& - \triangle \psi = - 4 \pi G \left(\rho - \rho_0 \right),
\label{hd1}
\end{eqnarray}
where $\rho$ is the density, $\uu$ -- velocity, $p$ -- pressure, 
$ \psi $ -- gravitational potential,
${\gg}=- \nabla \psi $ -- gravitational acceleration, 
$G$ -- gravitational constant, 
${\bm d} \equiv  (\zeta + {\mu}/3) \nabla (\nabla \cdot \uu) + \mu \Delta \uu $ stands for the viscous terms ($\zeta$ and $\mu$ being respectively the bulk and dynamic viscosity which are constants under isothermal assumption) and 
${\bm f}$ -- random stationary homogeneous external force. The system is closed with an 
isothermal equation of state $p=  c_s^2 \rho $, where $c_s$ is the Newtonian sound speed. In the Poisson equation (\ref{hd1}), we use the density fluctuation with respect to the spatial average $\rho_0 $ rather than the local density $\rho$, assuming 
periodic boundary conditions often used in simulations of turbulence in astrophysical systems.
The total energy density at any point of the flow field is the sum of kinetic, thermodynamic and gravitational potential energy densities at that point and can thus be written as 
\begin{equation}
{\pazocal E} =    \rho \bm u^2/2 + \rho e - \alpha \bm g^2/2 , \label{consE}
\end{equation}
where $\alpha  = 1/(4 \pi G)$ and $e\equiv c_s^2\ln \left(\rho/ \rho_0 \right)$ is the isothermal thermodynamic energy per unit mass. The minus sign accounts for the fact that gravitational potential energy is always due to attractive interactions. 
Now for an isothermal fluid 
\begin{equation}
\nabla p/\rho= c_s^2 \nabla \rho/\rho = c_s^2 \nabla ( \ln \rho) = \nabla e.  \label{simpli}
\end{equation} 
The second law of thermodynamics states that on a closed system (where the mass is conserved), the work done, which will be stored in the system as potential energy, can be obtained from the change in temperature (the internal energy) and from the change in entropy.  Unlike non-isothermal polytropic flows, for isothermal system, the thermodynamic energy is not contributed by the internal energy as the temperature is constant. 
Using equation (\ref{simpli}), the equation (\ref{momentum}) can therefore be rewritten as 
\begin{equation}
\partial_t \uu + (\uu \cdot \Nabla) \uu = - \nabla e + \gg + \rho^{-1} ({\bm d} + {\bm f}), \label{internal} .
\end{equation}

In order to derive the evolution equation for $\gg$, we take partial time derivative to both sides of the momentum conservation equation and then, using the continuity equation, we get 
\begin{equation}
\Nabla \cdot   {\partial_t \gg}  =  4 \pi G \nabla \cdot (\rho \uu) =>  \partial_t \gg - 4 \pi G \rho\uu = \nabla \times {\ag},  \label{evolution}
\end{equation}
where $ \ag $ is the vector potential related to gravitation. 
To show the conservation of energy when the viscous and forcing terms are neglected, 
we evaluate the time derivative of each term of total energy. We have (assuming periodic boundary conditions or zero velocity on the boundary surface)
\begin{eqnarray}
\frac{1}{2}  \int_V {\partial_t} (\rho \uu^2) d\bm x   &=& \int_V \left( \rho \uu \cdot \gg - \uu \cdot \nabla p \right)  d\bm x,  \\
\int_V {\partial_t (\rho e)} d\bm x &=& \int_V   \uu \cdot \nabla  p\, d\bm x, \\
- \frac{\alpha }{2}\int_V {\partial_t} \gg^2 d\bm x  &=& \int_V \left[ \nabla\! \cdot\! \left( \gg\! \times\! {\bm a_G} \right) - \rho \uu\! \cdot\! \gg\right]  d\bm x, 
\end{eqnarray}
where ${\bm  a}_G = \alpha{\bm A}_G/2$.
Adding up the three above expressions and assuming additionally that the boundary surface is either periodic or gravitationally equipotential ($\gg=-\nabla \psi = 0$), 
we can prove the total energy conservation for a non-viscous system without external forcing.


\section{Construction of correlators}\label{correl}
Although, it is the average total energy which is shown to be a constant in volume, we are basically interested in the behaviour of the turbulent energy. In order to understand turbulent energy, we  decompose every scalar and vector field at every point in space as a summation of the statistical average and the fluctuation field (i.e. Reynolds decomposition). So we have
\begin{align}
\rho &= \left\langle{\rho}\right\rangle + \tilde{\rho},  \ \    e =  \left\langle{e}\right\rangle + \tilde{e}, \ \ \uu =  \left\langle{\uu}\right\rangle + \tilde{\uu} \, \, \text{and}  \, \ \ \rho \uu = \left\langle{ \rho \uu}\right\rangle + \widetilde{\rho \uu} ,
\end{align}
where $\left\langle \cdot \right\rangle$ denotes the statistical average and the $\tilde{\xi}$ denotes the fluctuation of $\xi$ . Since we are assuming homogeneous turbulence, the statistical average is equivalent to the spatial average. 

For an incompressible system, the only component, which is called the average turbulent kinetic energy, is traditionally defined as the average of the square of the velocity fluctuations with respect to the statistical average of velocity \citep{Karman38, Pope2000}. For a compressible system, we do not have any strict recipe to define turbulent energy. One can then define the turbulent energy density to be 
\begin{equation}
{\pazocal E}_f =   \frac{\widetilde{ \rho \uu} \cdot  {\tilde \uu}}{2}  +  \tilde{\rho} \tilde{e} - \frac{\alpha}{2} \tilde{\bm g}^2 .
\end{equation}
According to the above definition, turbulent energy is purely produced by the field fluctuations. In case of compressible turbulence, this definition causes a problem. Like incompressible Navier-Stokes equations, compressible Navier-Stokes equations also remain invariant under Galilean transformation. Unlike the mean velocity field, this transformation cannot eliminate the mean density and mean thermodynamic energy field. In other words, the fluctuating total energy is not an inviscid invariant of compressible Navier-Stokes equations. For a quantity, which is not an inviscid invariant of the flow, the scaling relation and the plausibility of a cascade is not {obvious}. Hence, we define the turbulent energy density as the total energy density perceived with respect to a reference frame at which the mean velocity is zero. One can {readily} show that in such a reference frame, the velocity field is given by the fluctuations, whilst the density and the thermodynamic energy fields are expressed as the total value (mean + fluctuation). Just by observation, it is also easy to understand that under Galilean transformation, the gravitational acceleration $\gg$ also behaves like density and thermodynamic energy. {Thus}, the average total energy 
\begin{equation}
\left\langle {\pazocal E} \right\rangle =   \left\langle \frac{ \rho \tilde{  \uu} \cdot  {\tilde \uu}}{2}  +  {\rho} {e} - \frac{\alpha}{2} {\bm g}^2 \right\rangle =  \left\langle {\pazocal {E_H}} - \frac{\alpha}{2} {\bm g}^2   \right\rangle          , \label{consEb}
\end{equation}
is an inviscid invariant of {self-gravitating} compressible isothermal turbulence where $ {\pazocal {E_H}}$ is the hydrodynamic part. So, in the current case, the turbulent energy is the total energy of the flow with respect to a reference frame where the mean velocity of the flow is zero. The next step is the construction of the correlators which can be done simply by defining the symmetric two-point correlation functions as (note that $\uu \equiv \tilde{\uu}$ as $\left\langle \uu \right\rangle = 0$)
\begin{equation}
{\pazocal R}(\rr) = \left\langle \frac{R_\pazocal{E} + R'_\pazocal{E}}{2} \right\rangle \label{14}
\end{equation}
with 
\begin{align}
R_{\pazocal E} &\equiv  \frac{1}{2} \jm \cdot {\uu}'   + {\rho} {e}'  -  \frac{\alpha}{2} {\gg} \cdot {\gg}'  , \\
R'_{\pazocal E} &\equiv  \frac{1}{2} \jm'  \cdot {\uu}  +  {\rho}' {e}  -  \frac{\alpha}{2} {\gg}' \cdot {\gg} ,
\label{re}
\end{align} 
where the unprimed and the primed quantities correspond to the variables of the points situated at $\xx$ and $(\xx + \rr)$, respectively and $\jm \equiv \rho \uu$. It is noteworthy that the definition of kinetic energy correlator is not unique in compressible case. Our choice is particularly inspired by the nonlinear term of Navier-Stokes equation which is written as $\nabla \cdot (\jm \otimes \uu)$ where both {$\jm$} and $\uu$ carry physical meaning (momentum and velocity). The same form has also been used in Refs.~\cite{Graham10,Galtier11,Salvesen14}. Another possibility is to express the kinetic energy density as the square of the variable $\bm w \equiv \sqrt{\rho} \uu$ and to define the correlator as $\left\langle  \bm w \cdot \bm w' \right\rangle / 2$ in order to guarantee the positive definiteness of kinetic energy density in the spectral space \cite{Kida90, Miura95}. 	However, $\sqrt{\rho} \uu$ does not have clear physical meaning.  In fact, unlike incompressible case, in compressible turbulence, one should, in general, investigate the cospectral density of various pairs of physically meaningful variables, e.g., the momentum and the velocity, and \textit{a priori} there should be no constraint on the positivity of the cospectral density.

However, we must be careful to make sure it satisfies certain restrictions. In the single-point limit ${\pazocal R}(0)=\langle {\pazocal E} \rangle $, but this is not the only constraint that matters. To understand that, we write the the correlation function as a linear combination of correlators of kinetic energy (${\pazocal R}_K$), gravitational potential energy (${\pazocal R}_W$) and compressive potential energy (${\pazocal R}_U$) as
\begin{equation}
{\pazocal R}(\rr) = {\pazocal R}_K(\rr) + {\pazocal R}_U(\rr) + {\pazocal R}_W(\rr) .
\end{equation}
Let's concentrate on the compressive energy part ${\pazocal R}_U(\rr)=\langle {\rho} {e}'+ {\rho}' {e} \rangle/2$ (the gravitational and kinetic parts are already well defined). One can easily show that our ${\pazocal R}_U(\rr)$ is a particular case of a more general definition ${\pazocal R}_U(n;\rr)=(n-1)\langle {\rho} {e} \rangle/n+\langle {\rho} {e}'+ {\rho}' {e}\rangle/2n$, corresponding to $n=1$. Note that the single-point limit is still the same at any $n\ne0$, ${\pazocal R}_U(n;0)=\langle {\rho} {e} \rangle $, but the scale-dependent part $\langle {\rho} {e}' + {\rho}' {e}\rangle/2n$ vanishes at $n\rightarrow\infty$. We need to find a way to restrict $n$ based on the structure of NS equations or using some relevant exact solution. Since ${\pazocal R}_U(n;\rr)$ is ultimately a linear combination of single-point and two-point constituents, we will look for the value of $n$ that would give us the correct density of sound energy. This known linear solution will be sufficient to constrain the linear combination. 
One can {readily} show, for an isothermal or polytropic fluid, {that} the total energy density of the acoustic mode (${\pazocal E}_s$) is given by ({ignoring} gravity)
\begin{equation}
{\pazocal E}_s =   \frac{1}{2} \rho_0 \bm u^2 +  \frac{c_s^2}{2 \rho_0} {\rho}^2 , \label{consEa}
\end{equation}
whence the corresponding energy correlation function can be written as 
\begin{equation}
{\pazocal R}_s(\rr)=\frac{1}{2}\rho_0\langle\uu\cdot\uu'\rangle+\frac{c_{\rm s}^2}{2\rho_0}\langle{\rho}{\rho}'\rangle
\end{equation}
so that ${\pazocal R}_s(0)=\langle{\pazocal E}_s\rangle$, ${\pazocal R}_s(\infty)=0$, and the Fourier transform $\widehat{{\pazocal R}_s}(\kk)$ would give us the energy spectral density of sound.
In the acoustic limit, the kinetic and the potential energy follow detailed equipartition \citep{Sarkar, Sagaut}. The equipartition is also supported by recent numerical work \citep{KritsukFalko}. With our old expression of thermodynamic energy correlator $i.e.$ at $n=1$ \citep{Galtier11}, we get at acoustic limit, ${\pazocal R}_U(1;\rr)=\frac{c_{\rm s}^2}{\rho_0} \left(\langle\rho\rho'\rangle-\langle\rho^2\rangle/2 \right)$. After Fourier transform, it gives thermodynamic energy spectral density which is two times the required energy spectral density. This problem can be solved by choosing $n=2$ whence ${\pazocal R}_U(2;\rr)=\frac{c_{\rm s}^2}{2\rho_0}\langle{\rho}{\rho}'\rangle$, as needed. In that case, the general thermodynamic potential energy correlator looks like
\begin{equation}
{\pazocal R}_U(2;\rr)=  \frac{\langle {\rho} e\rangle}{2}+ \frac{\langle {\rho} {e}'+ {\rho}' {e} \rangle}{4}. \label{n2}
\end{equation}
Taking the new modifications into account, the hydrodynamic energy correlation functions can be expressed as 
\begin{equation}
R_{\pazocal H} \equiv (\jm \cdot {\uu}' + {\rho} {e}' + \rho e )/2,  \qquad  R'_{\pazocal H} \equiv ( \jm' \cdot {\uu} + {\rho}' {e} + \rho'e')/2,
\end{equation}
and therefore the total energy correlation (including the gravitational energy) can be {defined by}
\begin{equation}
R_{\pazocal E} \equiv (\jm  \cdot {\uu}' + {\rho} {e}' + \rho e - \alpha {\gg} \cdot {\gg}')/2,  \qquad  R'_{\pazocal E} \equiv (\jm' \cdot {\uu} + {\rho}' {e} + \rho'e' - \alpha {\gg}' \cdot {\gg})/2. \label{ecf}
\end{equation}

\section{Derivation of the exact relation}\label{exact}
The next step is to derive the evolution equations for ${\pazocal R}(\bm r)$. Using equations (\ref{continuity}), (\ref{momentum}), 
(\ref{hd1}), (\ref{internal}) and (\ref{evolution}), we can write (without the forcing and the {viscous} terms)
\begin{align}
\p_t \left\langle \jm \cdot \uu' \right\rangle &=   \langle \jm \cdot \p_t \uu' + \uu' \cdot \p_t \jm \rangle \nonumber\\
&= - \left\langle  \jm \cdot \left[  (\uu' \cdot \Nabla') \uu' + \Nabla' e' - \gg'  \right] + \uu' \cdot \left[  \Nabla \cdot  (\jm \otimes \uu) + \Nabla p - \rho \gg  \right]  \right\rangle  \nonumber\\
&= - \left\langle  \jm \cdot \left[  (\uu' \cdot \Nabla') \uu' + \Nabla' e' - \gg'  \right] + \uu' \cdot \left[  \uu (\Nabla \cdot \jm) + (\jm \cdot \nabla) \uu + \Nabla p - \rho \gg  \right]  \right\rangle  \nonumber\\
&= - \left\langle  \jm\cdot\Nabla'e' -p \theta'+ \jm \cdot\left[(\uu' \cdot \Nabla') \uu' -\gg'\right]+ \uu' \cdot \left[  \uu (\Nabla \cdot \jm) + (\jm \cdot \nabla) \uu - \rho \gg  \right]  \right\rangle , \\
\p_t \langle {\rho} {e}' \rangle &= \langle \rho \ \p_t e' + e' \ \p_t \rho \rangle \nonumber\\
&= - \left\langle \Nabla \cdot ( \jm e' ) + \Nabla' \cdot ( \jm' e') -   \rho e'\theta'  +  p \theta'  \right\rangle   \nonumber\\
&= - \left\langle -\jm\cdot\Nabla'e' +  p \theta'+ \rho\left[\Nabla' \cdot (  \uu' e') -  e'\theta' \right] \right\rangle   \nonumber\\
&= - \left\langle -\jm\cdot\Nabla'e' +  p \theta' + \rho\uu'\cdot\Nabla'e'\right\rangle   \\
\p_t \langle \rho e \rangle &= \p_t \langle \rho' e' \rangle = \langle \rho \ \p_t e + e \ \p_t \rho \rangle  \nonumber\\
&= - \left\langle p \theta \right\rangle  \\
\p_t \left\langle {\gg} \cdot {\gg}' \right\rangle &= \p_t \left\langle {\gg} \cdot {\gg}' \right\rangle - \p_t \left( \overline{\gg} \cdot \overline{\gg} \right) = \langle  \gg \cdot \p_t \gg' + \gg' \cdot \p_t  \gg \rangle \nonumber\\
&= \left\langle \gg' \cdot \left[  4 \pi G \jm   +  \Nabla \times \ag \right] + \gg \cdot \left[  4 \pi G \jm'   +  \Nabla' \times \ag' \right] \right\rangle \nonumber\\
&=  \langle \Nabla \cdot (\ag \times \gg') + \Nabla' (\ag' \times \gg) \rangle + 4 \pi G \langle \jm \cdot  \gg'  + \jm'  \cdot  \gg  \rangle \nonumber\\
&=  \alpha^{-1} \left\langle \jm \cdot  \gg'  + \jm'  \cdot  \gg  \right\rangle   ,
\end{align}
where $ \theta \equiv \nabla \cdot \uu$ and  we have used the statistical homogeneity to obtain $ \langle  \Nabla \cdot (\ag \times \gg') \rangle = - \langle  \Nabla' \cdot (\ag \times \gg')\rangle = \langle \ag \cdot (\Nabla' \times \gg') \rangle =0 $. In the following, we derive two {different} forms of the exact relation. 

\subsection{\textbf{In terms of two-point differences}}

This form looks nearly similar to the previously derived exact relations for compressible turbulence \citep{Galtier11, Banerjee13, Banerjee14}. In order to obtain the concerned form, we combine the above expressions and obtain for $R_{\pazocal H}$ (and similarly for $R'_{\pazocal H}$) 
\begin{align}
\frac{\partial \left\langle R_{\pazocal H} \right\rangle }{\partial t}  &=  - \Nabla_\rr  \cdot  \left\langle \frac{1}{2}\left[ \jm \cdot \uu' + \rho e' \right]\delta \uu + \frac{\jm e'}{2}  \right\rangle +  \frac{1}{2} \left\langle \theta' \left( \jm \cdot \uu' + \rho e' \right) -  p \theta + \jm \cdot \gg' + \uu' \cdot \rho \gg  \right\rangle  ,
\\
\frac{\partial \left\langle R'_{\pazocal H} \right\rangle }{\partial t}  &=  - \Nabla_\rr  \cdot  \left\langle  \frac{1}{2}\left[ \jm' \cdot \uu + \rho' e \right]  \delta \uu - \frac{\jm' e}{2}   \right\rangle + \frac{1}{2} \left\langle \theta  \left( \jm' \cdot \uu + \rho' e  \right) - p' \theta' + \jm' \cdot \gg + \uu \cdot \rho' \gg' \right\rangle .
\end{align}

After {some} manipulations and reinserting the forcing and {viscous} terms, we get
\begin{align}
& \frac{\partial \left\langle R_{\pazocal E} + R'_{\pazocal E} \right\rangle }{\partial t}  \nonumber \\
& = - \Nabla_\rr \cdot \left\langle \left( R_{\pazocal H} + R'_{\pazocal H} \right) \delta \uu \right\rangle \nonumber + \frac{1}{2} \left\langle \rho \gg \cdot \uu' + \rho' \gg' \cdot \uu - \jm \cdot \gg' - \jm' \cdot \gg \right\rangle + 2{\pazocal D} + 2{\pazocal F}  \nonumber  \\
&+ \left\langle \theta  \left[ R'_{\pazocal H} - \frac{p}{2} \right] + \theta'  \left[ R_{\pazocal H}  - \frac{p'}{2} \right] - \frac{1}{2} \left(  \jm \cdot \nabla' e' + \jm ' \cdot \nabla e  \right)  \right\rangle 
\nonumber  \\
&= \frac{1}{2} \Nabla_\rr \cdot \left\langle \left[ {\delta \jm \cdot \delta \uu } + \delta \rho \delta e  \right] \delta \uu  \right\rangle   
+ \frac{1}{2} \left\langle \delta \jm \cdot \delta \gg - \delta (\rho \gg ) \cdot \delta \uu \right\rangle + 2{\pazocal D} + 2{\pazocal F} \nonumber  \\
&+ \left\langle \theta  \left( R'_{\pazocal H} - {\pazocal E}'_{\pazocal H}  \right) + \theta'  \left( R_{\pazocal H} - {\pazocal E}_{\pazocal H}    \right) \right\rangle +  \frac{1}{2}\left\langle \left( \Nabla \cdot \jm \right)e' + \left( \Nabla' \cdot \jm' \right)e   -   \left( \theta p + \theta'  p' \right) \right\rangle  \nonumber \\
&=  \frac{1}{2} \Nabla_\rr \cdot \left\langle \left[ {\delta \jm \cdot \delta \uu } + \delta \rho \delta e  \right] \delta \uu  \right\rangle + \frac{1}{2} \left\langle \delta \jm \cdot \delta \gg - \delta (\rho \gg ) \cdot \delta \uu \right\rangle + 2{\pazocal D} + 2{\pazocal F} \nonumber  \\
&+ \frac{1}{2} \left\langle \delta \theta \delta \left( \jm \cdot \uu + \rho e \right) - \delta \left( \theta \uu \right) \cdot \delta \jm - \delta \left( \theta e \right) \cdot \delta \rho +  \delta \jm \cdot \delta \left( \nabla e \right) \right\rangle  - \left\langle \nabla \cdot \left( p \uu \right) \right\rangle,  \label{premain}
\end{align}
where for any variable $X$, $\delta X \equiv X (\xx + \rr) - X (\xx)=X'-X$ and 
${\pazocal D}$ and ${\pazocal F}$ denote the {viscous} contribution and the forcing term, respectively. The algebraic expressions for ${\pazocal D}$ and ${\pazocal F}$ are given as 
\begin{align}
{\pazocal D} &= \frac{1}{4}\left\langle \jm \cdot {\bm d'}/\rho' + \uu' \cdot {\bm d} + \jm' \cdot {\bm d}/\rho + \uu \cdot {\bm d'} \right\rangle,\label{diss} \\
{\pazocal F} &= \frac{1}{4}\left\langle \jm \cdot {\bm f'}/\rho' + \uu' \cdot {\bm f} + \jm' \cdot {\bm f}/\rho + \uu \cdot {\bm f'} \right\rangle.\label{for}  
\end{align}
In addition, to obtain the final step, we used the relation
\begin{equation}
\left( R_{\pazocal H} + R'_{\pazocal H}  \right) - \left( {\pazocal E}'_{\pazocal H} + {\pazocal E}_{\pazocal H} \right) =  - \frac{1}{2} \left( {\delta \jm \cdot \delta \uu } + \delta \rho \delta e \right).
\end{equation}

Now for a state, where the average energy and its correlator vanish by the balance between forcing and the viscous dissipation (weaker approximation than a stationary state), the left hand side of equation (\ref{premain}) vanishes. In addition, we concentrate on the inertial zone 
(which is assumed to exist for compressible turbulence \citep{Aluie11,Aluie13}), where the {viscous} effects can be neglected and the external forcing is assumed to be the only source of the energy input 
\begin{equation}
- 4 \varepsilon =   \Nabla_\rr \cdot \left\langle \left( {\delta \jm \cdot \delta \uu } + \delta \rho \delta e  \right) \delta \uu  \right\rangle
+  \left\langle   \delta\left( \jm \cdot \uu + \rho e \right)\delta \theta + \delta \jm \cdot \delta  \left( \nabla e  -   \uu \theta \right)  - \delta \rho \delta \left(e \theta  \right) \right\rangle  +  \left\langle \delta \jm \cdot \delta \gg - \delta \uu\cdot\delta (\rho \gg )  \right\rangle,   \label{result1}
\end{equation}
where $\varepsilon = {\pazocal F}$ represents the {(generally, scale-dependent)}  mean rate of energy injection by the external force and the term $ \left\langle \nabla \cdot \left( p \uu \right) \right\rangle $ vanishes by the application of Gauss' divergence theorem.  

Equation (\ref{result1}) is the main result of the paper. This is an exact relation for three-dimensional, homogeneous, isothermal self-gravitating turbulence valid in the limit of asymptotically large Reynolds numbers. Unlike the {relationships} obtained in previous papers \citep{Galtier11, Banerjee13}, {the} current {form} is expressed {solely} in terms of two-point differences. Due to the modification in the definition of the correlation functions, the nongravitational {part} of (\ref{result1}) {differs} from that obtained in \citet{Galtier11}.  Interestingly, one can show that the velocity--pressure-dilatation correlation does not appear in this exact relation which was claimed previously \citep{Aluie13}. 

We now rewrite (\ref{result1}) in a slightly different form
\begin{align}
- 4 \varepsilon = &  \Nabla_\rr \cdot \left\langle \left(\delta \jm \cdot \delta \uu\right) \delta \uu  \right\rangle 
+  \left\langle   \delta\left( \jm \cdot \uu\right)\delta \theta -   \delta \jm \cdot \delta  \left(\uu \theta \right)\right\rangle \nonumber \\
+ &\Nabla_\rr \cdot \left\langle \delta \rho \delta e  \delta \uu\right\rangle+ \left\langle \delta\left(\rho e \right)\delta \theta + \delta \jm \cdot \delta  \left( \nabla e\right)    - \delta \rho \delta \left(e \theta  \right) \right\rangle  \nonumber \\
+ & \left\langle \delta \jm \cdot \delta \gg - \delta \uu\cdot\delta (\rho \gg )  \right\rangle,   \label{result1a}
\end{align}
which is convenient for the discussion that follows. Here, the right-hand side (rhs) is divided across three lines based on the energies involved: kinetic, thermodynamic, gravitational, respectively. The $\Nabla_\rr \cdot \langle\ldots\rangle$ terms on lines 1 and 2 represent the divergence of kinetic and thermodynamic energy fluxes.
The {last line} of (\ref{result1a}) describes {a scale-dependent net energy input due to self-gravity}. Interestingly, this new contribution can neither be expressed as a flux term nor a usual source term. However, under the assumption of isotropy, the gravitational energy source in equation (\ref{premain}) can be written as
\begin{equation}
S(r)=\langle\rho\gg\cdot\uu' + \rho'\gg'\cdot\uu - \gg'\cdot\jm - \gg\cdot\jm'  \rangle/2 = \langle\rho\gg\cdot\uu' - \gg'\cdot\jm\rangle,
\end{equation}
Consider correlation lengths $L_{\rho}$, $L_g$, $L_u$ associated with the density, gravitational acceleration and velocity, respectively. In supersonic turbulence, density is very short-correlated compared to the velocity, $L_{\rho}\ll L_u$ \citep{kritsukjfm2013}. This is reflected, for instance, in shallow density spectra ($P(\rho,k)\propto k^{-1}$ at Mach 6 or even shallower at higher Mach numbers), while the velocity spectra scale $\propto k^{-2}$ \citep{kritsuk...07}. 
As one can expect that $L_{\rho}\sim L_g\ll L_u$ and so the density strongly correlates with the acceleration $\gg$, at $r\gtrsim L_u$ the first term will dominate as small-scale momentum decorrelates from small-scale acceleration
\begin{equation}
S(r)=\langle\rho\gg\cdot\uu' - \jm\cdot\gg'\rangle\approx\langle\rho\gg\cdot\uu\rangle - \langle\jm\rangle\cdot\langle\gg'\rangle\approx\langle\rho\gg\cdot\uu\rangle. 
\label{cor2}
\end{equation}
Case (\ref{cor2}) represents gravitational forcing of the turbulence at relatively large scales and thus $S(r)$ can be moved to the left-hand side of (\ref{result1}), which includes contributions from an external large-scale acceleration of the form similar to (\ref{cor2}). 

A special case when gravitational terms in (\ref{result1}) may cancel exactly relates to the so-called Zeldovich approximation \citet{zeldovich70} that assumes a predominantly potential velocity field ($\bm\nabla\bm\times\bm u=0$) such that $\bm u=\lambda\bm g$, where $\lambda$ is a constant \citep{shandarin.89}. Its generalization known as the {\em adhesion model} \citep{gurbatov.84,gurbatov..12} relies on multidimensional Burgers' equation in the limit of vanishing viscosity to describe structure formation in a uniform cold gravitationally unstable gas with random initial velocity perturbations \citep[e.g.][]{vergassola...94,ascasibar..13}. The cancellation stems from `minimization' of nonlinear terms, which implies enslaving of the velocity by the density through the gravitational potential. If relaxation (with quite different initial conditions in our case) would indeed favor depleted nonlinearity in some regime (e.g. in shock-compressed layers prone to collapse), the cancellation of gravitational terms would be selected naturally. The cancellation works whenever one of the eigenvalues of the rate-of-strain matrix $\partial_iu_j=\partial_i\partial_j\phi$ dominates over the other two; here $\phi$ is the velocity potential and $\bm u=-\nabla\phi$. This is the case for pancake-like structures (such as shock-compressed layers), but not for (cylinder-like) filaments in 3D, which have two dominating eigenvalues.

\subsection{\textbf{In terms of two-point correlation functions}}\label{too}
An alternative way to express the obtained exact relation is in terms of two-point correlation functions. Hereafter, we use the suffixes \textit{K}, \textit{U} and \textit{W} to denote quantities related to the kinetic, the thermodynamic and the gravitational energies, respectively. These correlation functions are grossly classified into three parts: (i) the transfer terms ${\pazocal T_{K, U, W}} $; (ii) the exchange (or cross-) terms ${\pazocal X}_{K\rightarrow U}$ and ${\pazocal X}_{K\rightarrow W}$; and (iii) the source term ${\pazocal S}_U $. The transfer terms represent the net energy correlation flux of a given type (kinetic, thermodynamic or gravitational) which gets transferred from one scale to other whereas the exchange terms represent the energy which gets converted to another form at a given scale $r$ (kinetic to thermodynamic and vice-versa, for example). The source term represents the single-point contribution in $\pazocal R_U(r)$, see (\ref{n2}). The exchange terms cancel out in the evolution equation for the correlator of the total energy $\pazocal R$, but appear in the evolution equations for ${\pazocal R}_K$, ${\pazocal R}_U$, and ${\pazocal R}_W$,
\begin{align}
\partial_t{\pazocal R}(\rr,t)&={\pazocal T}_K(\rr,t)+{\pazocal T}_{U}(\rr,t)+{\pazocal T}_{W}(\rr,t)+{\pazocal D}(\rr,t)+{\pazocal F}(\rr,t)+{\pazocal S}_U(t),\\
\partial_t{\pazocal R}_K(\rr,t)&={\pazocal T}_K(\rr,t)-{\pazocal X}_{K\rightarrow U}(\rr,t)-{\pazocal X}_{K\rightarrow W}(\rr,t)+{\pazocal D}(\rr,t)+{\pazocal F}(\rr,t),\\
\partial_t{\pazocal R}_U(\rr,t)&={\pazocal T}_U(\rr,t)+{\pazocal X}_{K\rightarrow U}(\rr,t)+{\pazocal S}_U(t),\\
\partial_t{\pazocal R}_W(\rr,t)&={\pazocal T}_W(\rr,t)+{\pazocal X}_{K\rightarrow W}(\rr,t),\label{potent}
\end{align}
where the various transfer, exchange, and source functions from the above conservation laws are defined as
\begin{align}
{\pazocal T}_K = &-\frac{1}{4} \left\langle \right. \jm \cdot  \left( \uu' \cdot \nabla' \right) \uu' 
+ \jm' \cdot  \left( \uu \cdot \nabla \right) \uu \nonumber\\
&+ \; \uu \cdot \left[ \left( \nabla' \cdot \jm' \right) \uu' + \left(\jm' \cdot \nabla' \right) \uu'  \right] 
+ \uu' \cdot \left[ \left( \nabla \cdot \jm \right) \uu + \left(\jm \cdot \nabla \right) \uu \right]   \left.\right\rangle, \\
{\pazocal X}_{K\rightarrow U} = &+\frac{1}{4} \left\langle \right. 
\jm \cdot \nabla' e' + 	\jm' \cdot \nabla e- p' \theta - p \theta'   \left.\right\rangle, \\
{\pazocal T}_{U} = & - \frac{1}{4}\left\langle \right.  \rho \uu' \cdot \nabla' e' + \rho' \uu \cdot \nabla e \rangle \rangle, \\
{\pazocal S}_{U} =  &+ \frac{1}{2}\langle p \theta \rangle, \\
{\pazocal T}_{W} = &- \frac{1}{4} \langle  \jm \cdot \gg' + \jm'  \cdot \gg - \rho\gg\cdot\uu' - \rho'\gg'\cdot\uu  \rangle,\label{tw}\\
{\pazocal X}_{K\rightarrow W} = &-\frac{1}{4} \left\langle \right. \rho\gg\cdot\uu' + \rho'\gg'\cdot\uu +  \jm \cdot \gg' + \jm'  \cdot \gg 
\left.\right\rangle . \label{xkw}
\end{align}
Now for a state where the time derivative of the average total energy correlator vanishes due to the balance between forcing and dissipation, we can write 
\begin{equation}
{\pazocal T}_K(\rr) + 	{\pazocal T}_{U}(\rr) + 	{\pazocal T}_{W}(\rr) +  {\pazocal S}_{U} (\rr) = - {\pazocal F}(\rr)-{\pazocal D}(\rr), \label{result2}
\end{equation}
For the so-called inertial zone, we can neglect the dissipative terms and finally have 
\begin{equation}
{\pazocal T}_K(\rr) + 	{\pazocal T}_{U}(\rr) + 	{\pazocal T}_{W}(\rr) +  {\pazocal S}_{U} (\rr)= - \varepsilon (\rr) , \label{result2a}
\end{equation}
which is equivalent to the equation (\ref{result1}).

We note by looking at (\ref{tw}) and (\ref{xkw}) that, unlike in other cases, the transfer and cross-terms in (\ref{potent}) partly overlap, that is, the evolution equation for ${\pazocal R}_W$ can be written in a very simple form with a symmetric part of the cross-covariance of $\jm$ and $\gg$ on the rhs,
\begin{equation}
\partial_t{\pazocal R}_W(\rr,t)=-\frac{1}{2} \left\langle  \jm \cdot \gg' + \jm'  \cdot \gg \right\rangle ,
\end{equation}
which would statistically vanish in stationary conditions.
At the same time, ${\pazocal X}_{W\rightarrow K}=-{\pazocal X}_{K\rightarrow W}$ describes nothing but the broadband gravitational forcing ${\pazocal F}_g(\rr)$, which has the same functional form as (\ref{for}). Hence, we can interpret the transfer of gravitational potential energy ${\pazocal T}_{W}(\rr)={\pazocal F}_g(\rr)+\partial_t{\pazocal R}_W(\rr)$ as a result of imbalance (across scales) between gravitational forcing and the rate of accumulation of potential energy. In stationary conditions (e.g., when gravity is uniformly bounded and turbulence is fully developed, one obtains $\left\langle  \jm \cdot \gg' + \jm'  \cdot \gg \right\rangle=0$ and thus ${\pazocal T}_{W}(\rr) = {\pazocal F}_g(\rr)$, so that the transfer function simply represents forcing due to self-gravity, see also (\ref{cor2}). If instead stationarity is broken (due, e.g., to an ongoing gravitational phase transition) one can think of a scenario discussed at the end of previous section, where the transfer would vanish due to the cancellation of gravitational terms on the rhs of (\ref{tw}). This would correspond to a balanced case of production and storage of the potential energy.

Now, focusing on a statistically stationary state, we can finally obtain
\begin{equation}
{\pazocal T}_K(\rr) + 	{\pazocal T}_{U}(\rr) + 	\widetilde{\pazocal T}_{W}(\rr) = - {\pazocal F}(\rr)-{\pazocal D}(\rr), \label{result2b}
\end{equation}
where we assumed that $\partial_t{\pazocal R}$, the source ${\pazocal S}_{U}=-\partial_t\langle\rho e\rangle /2 $, and $\partial_t{\pazocal R}_W$  statistically vanish, and introduced new notation for the stationary transfer function
\begin{equation}
\widetilde{\pazocal T}_{W} = \langle \rho\gg\cdot\uu' + \rho'\gg'\cdot\uu  \rangle/4.\label{tw1}
\end{equation}

\subsection{Energy transfer in spectral space}

We shall now use the correlation form (\ref{result2b}) of the exact relation to study the energy budget in spectral space. Taking Fourier transform of (\ref{result2}) and integrating over spherical shells in $k$-space, we obtain
\begin{equation}
	T(k)\equiv T_K(k)+T_U(k)+\widetilde{T}_W(k)=-F(k)-D(k),\label{dub}
\end{equation}
where $T(k)=\int\widehat{\pazocal T}(\bm \kappa)\delta(k-|\bm\kappa|)d\bm\kappa$ is the total energy spectral transfer function, $F(k)=\int\widehat{\pazocal F}(\bm \kappa)\delta(k-|\bm\kappa|)d\bm\kappa$ is the energy injection spectrum, and $D(k)=\int\widehat{\pazocal D}(\bm \kappa)\delta(k-|\bm\kappa|)d\bm\kappa$ is the energy dissipation (here $\widehat{\pazocal A}(\bm\kappa)$ denotes the Fourier transform of ${\pazocal A}(\rr)$). Finally, spectral energy fluxes through wavenumber $k$ can be defined in a standard form 
	\begin{equation}
	\Pi(k)\equiv\Pi_K(k)+\Pi_U(k)+\widetilde{\Pi}_W(k)=\int_k^{\infty}T(\kappa)d\kappa,
	\end{equation}
	or, equivalently,
	\begin{equation}
	T(k,t)=-\frac{\partial\Pi(k,t)}{\partial k}.
	\end{equation}
Under stationary conditions, steady energy injection by the forcing balances viscous dissipation, $\int_0^{\infty}[F(k)+D(k)]dk = 0$, which implies $\int_0^{\infty}T(k)dk=0$, and hence the flux can also be computed as $\Pi(k)=-\int_0^kT(\kappa)d\kappa$. In the limit of vanishing viscosity, in the inertial range, the spectral energy flux is constant, 
\begin{equation}
\Pi(k)={\epsilon}\equiv\int_0^{\infty}F(k)dk=\int\widehat{\pazocal F}(\bm\kappa)d\bm\kappa=\int\varepsilon(\rr)d\rr, \label{result3}
\end{equation}
and the spectral energy transfer function $T(k)=0$. Thus (\ref{result3}) is a spectral-space analogue of our main result (\ref{result1}).

Finally, we define relevant spectral energy densities similar to $F(k)$ above, namely, $K(k)\equiv\int\widehat{\pazocal R}_K(\bm \kappa)\delta(k-|\bm\kappa|)d\bm\kappa$, $U(k)\equiv\int\widehat{\pazocal R}_U(\bm \kappa)\delta(k-|\bm\kappa|)d\bm\kappa$, and $W(k)\equiv\int\widehat{\pazocal R}_W(\bm \kappa)\delta(k-|\bm\kappa|)d\bm\kappa$.
The Fourier transforms involved are real by construction of cross-covariances in (\ref{14}). 
Moreover, these (co)spectral densities satisfy Parseval's theorem \citep{kritsuk..17, KritsukFalko}.
 Also note that $K(0)=0$ if $\langle\uu\rangle=0$ and $W(0)=0$ since $\langle\rho\rangle=\rho_0$, but $U(0)=\langle\rho e\rangle/2$ according to (\ref{n2}). Since $\alpha \widehat{\left\langle \gg\cdot\gg'\right\rangle}=\alpha |\widehat{\gg}(\bm k)|^2=\alpha^{-1} k^{-2}|\widehat{\rho}(\bm k)|^2$, the gravitational energy spectral density can be readily expressed in terms of the density power spectrum as $W(k)=-2\pi Gk^{-2}\int |\widehat{\rho(\bm \kappa)}|^2\delta(k-|\bm\kappa|)d\bm\kappa$.

\section{Summary and Discussion}
In this work, we have clarified the definition of turbulent energy as it is used in the framework of the derivation of exact relations for homogeneous turbulence. With respect to the previously derived relations of compressible hydrodynamic and magnetohydrodynamic turbulence \citep{Galtier11, Banerjee13, Banerjee14}, here we have modified the definition of the two-point energy correlation function by putting an additional constraint of detailed equipartition (in spectral space) between dilatational kinetic and thermodynamic potential energy in the acoustic limit. Using the modified correlation function, we showed that the resulting relation is indeed much simpler and is free from any pressure-dilatation--velocity correlation. In addition, both flux and source terms are now expressed in terms of two-point fluctuations. We also derived an alternative form of the relation formulated solely in terms of correlation functions. We then used the correlation form to derive the scale-by-scale energy budget in spectral space. The current work includes the effects of self-gravity which, presumably, contributes as a forcing term acting across scales in statistically stationary conditions. However, the gravitational part does not contribute to the two-point energy flux. The total gravitational contribution is shown to vanish under Zeldovich approximation. 

The obtained exact relations can be verified using numerical data for both subsonic and supersonic isothermal turbulence. The present work can be generalized in the case of compressible magnetohydrodynamic turbulence. Another non-trivial extension will be to use non-isothermal closures. 

\begin{acknowledgments}
We thank S\'ebastien Galtier  and Gregory Falkovich for stimulating discussions.
The work of A.K. was supported in part by the National Science Foundation Grant No.~AST-141227.
\end{acknowledgments}


\end{document}